\documentclass[prl,twocolumn,showpacs,amsmath,amssymb,epsfig]{revtex4-1} 
 
\usepackage{graphicx}
\usepackage{dcolumn}
\usepackage{bm}
\usepackage{mathrsfs}

\usepackage{natbib}
\usepackage[text={7in,9.5in},centering]{geometry}

\usepackage{hyperref}
\usepackage{url}

\usepackage{epsfig}
\usepackage{amsmath}
\usepackage{amssymb}
\usepackage{textcomp}

\usepackage{color}
\usepackage{float}

\usepackage{stackengine}
\usepackage{verbatim}
\begin{document}

\title{New anomaly observed in $^{12}$C supports the existence and the
  vector character of the hypothetical X17 boson}

\author{A.J. Krasznahorkay}
\email{kraszna@atomki.hu}
\author{A.~Krasznahorkay}
\altaffiliation{Currently working at CERN, Geneva, Switzerland}
\author{M. Begala}
\author{M. Csatl\'os}
\author{L. Csige}
\author{J. Guly\'as}
\author{A. Krak\'o}
\author{J. Tim\'ar}
\author{I. Rajta}
\author{I. Vajda}
\affiliation{Institute for Nuclear Research  (ATOMKI),
  P.O. Box 51, H-4001 Debrecen, Hungary}
\author{N.J. Sas}
\affiliation{University of Debrecen, 4010 Debrecen, PO Box 105, Hungary}

%%%%%%%%%%%%%%%%%%%%%%%%%%%%%%%%%%%%%%%%%%%%%%%%%%%%%%%%%%%%%%%%%%%%%%%
\begin{abstract}
Employing the $^{11}$B(p,$\gamma$)$^{12}$C nuclear reaction, the angular
correlation of $e^+e^-$ pairs was investigated in the angular range of
$40^\circ\Theta\leq 175^\circ$ for five
different proton energies between E$_p$ = 1.50 - 2.5~MeV. At
small angles ($\Theta\leq 120^\circ$), the results can be well
interpreted by the internal pair creation process of electromagnetic
radiations with E1 and M1 multipolarities and by the external pair
creation in the target backing. 
However, at angles greater than $120^\circ$, additional
count excesses and anomalies were observed, 
which could be well accounted for by the existence of the previously
suggested hypothetical X17 particle.
Our results suggest that the X17 particle was generated mainly in E1
radiation. The derived mass of the particle is
$m_\mathrm{X}c^2$=17.03$\pm 0.11 (stat) \pm 0.20 (syst)$~MeV.
According to the mass, and to the derived branching ratio
($B_x=3.6(3)\times10^{-6}$), this is likely the same X17 particle that we
recently suggested for describing the anomaly observed in the decay of
$^8$Be and $^4$He.

%%%%%%%%%%%%%%%%%%%%%%%%%%%%%%%%%%%%%%%%%%%%%%%%%%%%%%%%%%%%%%%%%%%%%%%%%  
\end{abstract}


\maketitle

\section{Introduction}

Very challenging nuclear physics experiments were initiated in 1978 to
detect a new particle, the axion, predicted by Weinberg
\cite{we78,wi78}.  The quantum chromodynamics axion is one of the most
compelling solutions to the strong CP (charge conjugation parity
symmetry) problem. Donelly \cite{do78} proposed to study the angular
correlation of the $e^+e^-$ pairs created in $1^+\rightarrow 0^+$
nuclear transitions as a signature for the decay of the
axion. However, it was quickly ruled out in the MeV/c$^2$ mass
regime.

The later introduced dark photon \cite{ga84,ho86}
is proposed as
a force carrier connected to dark matter. In a minimal scenario, this
new force can be introduced by extending the gauge group of the
Standard Model  with a new abelian U(1) gauge symmetry.

Pierre Fayet suggested a generalized dark photon model that would
produce a light gauge boson with an extra U(1) gauge group  already in
1980 \cite{Fayet:1980ad, Fayet:1980rr, Fayet:1980ss}. Such a generalized
dark photon may act as a mediator of light dark matter annihilations,
possibly allowing for lighter than GeV/c$^2$ dark matter particles
\cite{Boehm:2003hm, Fayet:2004bw}. More recent discussions on the light
U boson as the mediator of a new force, coupled to a combination of
Q, B, L and dark matter can be found in Ref~\cite{fa17,fa21}.

Searches for light particles, especially a 9~MeV/c$^2$
particle suggested by Fokke de Boer and coworkers, have
been performed in Frankfurt \cite{bo01}. These results, although
with little confidence, appeared to confirm the existence of
the 9~MeV/c$^2$ particle. However, due to the closure of the
accelerator, these experiments could not be continued. 

Recently, we studied electron-positron angular correlations for the
17.6~MeV and 18.15~MeV transitions in $^8$Be and an anomalous angular
correlation was observed for the 18.15~MeV transition \cite{kr16}.
This was interpreted as the creation and decay of an intermediate
bosonic particle with a mass of
$m_\mathrm{X}c^2$=16.70$\pm$0.35(stat)$\pm$0.5(sys)~MeV, which is now
called X17. The possible relation of the X17 boson to the dark matter
problem triggered an enormous interest in the wider physics community
\cite{ins,ag21}.

The first theoretical interpretation of the experimental results was
performed by Feng et al. \cite{fe16,fe17}.  They explained the anomaly
with a 16.7 MeV,  vector gauge boson X17, which may
mediate a fifth fundamental force with some coupling to Standard Model
(SM) particles. The theory of the dark photon has been generalized to
the fact that the new particle is coupled not only to electric charges
but also to quarks.

Constraints on the coupling constants of such a new particle, notably
from searches for $\pi_0\rightarrow Z' + \gamma$ by the NA48/2
experiment \cite{ba15} was also taken into account by Feng and
co-workers \cite{fe16,fe17}. Based on their results, the X17 particle
couples much more strongly to neutrons than to protons, so the
particle was named protophobic.

Zhang and Miller \cite{zh17} investigated the nuclear transition form
factor as a possible origin of the anomaly, but they found the
concluded form factor unrealistic for the $^8$Be nucleus.

Ellwanger and Moretti suggested another interpretation of the
experimental results in view of a light, pseudoscalar particle
\cite{ell16}. They
predicted about ten times smaller branching ratio in case of the 17.6
MeV transition compared to the 18.15 MeV one, which is in nice
agreement with our results.

Subsequently, many studies with different models have been
performed including an extended two Higgs doublet model \cite{de17,de19,ro19}.
They showed that the anomaly can be
described with a very light Z$_0$ bosonic state, with significant axial
couplings.  

In parallel to these recent theoretical studies, we re-investigated
the $^8$Be anomaly with an improved experimental setup \cite{kra17,kra19,kr17}. 

Recently, we also observed a similar anomaly in $^4$He
\cite{kra19,kr21}. The signal could be described by the creation and
subsequent decay of a light particle during the proton capture process
on $^3$H to the ground state of the $^{4}$He nucleus. The derived mass
of the particle ($m_\mathrm{X}c^2 = 16.94 \pm 0.12$(stat.)$\pm
0.21$(syst.)~MeV) agreed well with that of the proposed X17
particle. It was also shown, that the branching ratios of the X17
particle compared to the $\gamma$-decay
are identical within uncertainties for three beam energies,
proving that the X17 particle was most likely formed in direct proton
capture, which has a dominant multipolarity of E1.  Our results
obtained for $^4$He at different beam energies agree well with the
present theoretical results of Viviani et al. \cite{vi21}.

Referring to our manuscript \cite{kr21}, Feng and co-workers have
communicated a work very recently with the title of ,,Dynamical
evidence for a fifth force explanation of the ATOMKI nuclear
anomalies'' \cite{fe20}, in which they propose to study the E1 ground
state decay of the 17.2 MeV J$^\pi$ = 1$^-$ state in $^{12}$C in order
to determine if X17 has a vector or axial-vector character.

In the present work, we investigated the 17.2 MeV $1^-\rightarrow 0^+$
transition of $^{12}$C to search for signatures of the creation of the
X17 particle.

\section{Experimental methods}

The experiments were performed in Debrecen, Hungary at the 2 MV Tandetron
accelerator of ATOMKI. The E$_x$=17.23 MeV (J$^\pi$ = 1$^-$)
\cite{aj90} state of $^{12}$C was excited by the
$^{11}$B(p,$\gamma$)$^{12}$C nuclear reaction.  Due to its large cross
section, the reaction is also widely used for detector
calibrations. The resonant proton energy is E$_p$ = 1.388 MeV
\cite{aj90,se65}.

\begin{figure}[htb]
  \begin{center}
    \includegraphics[scale=0.40]{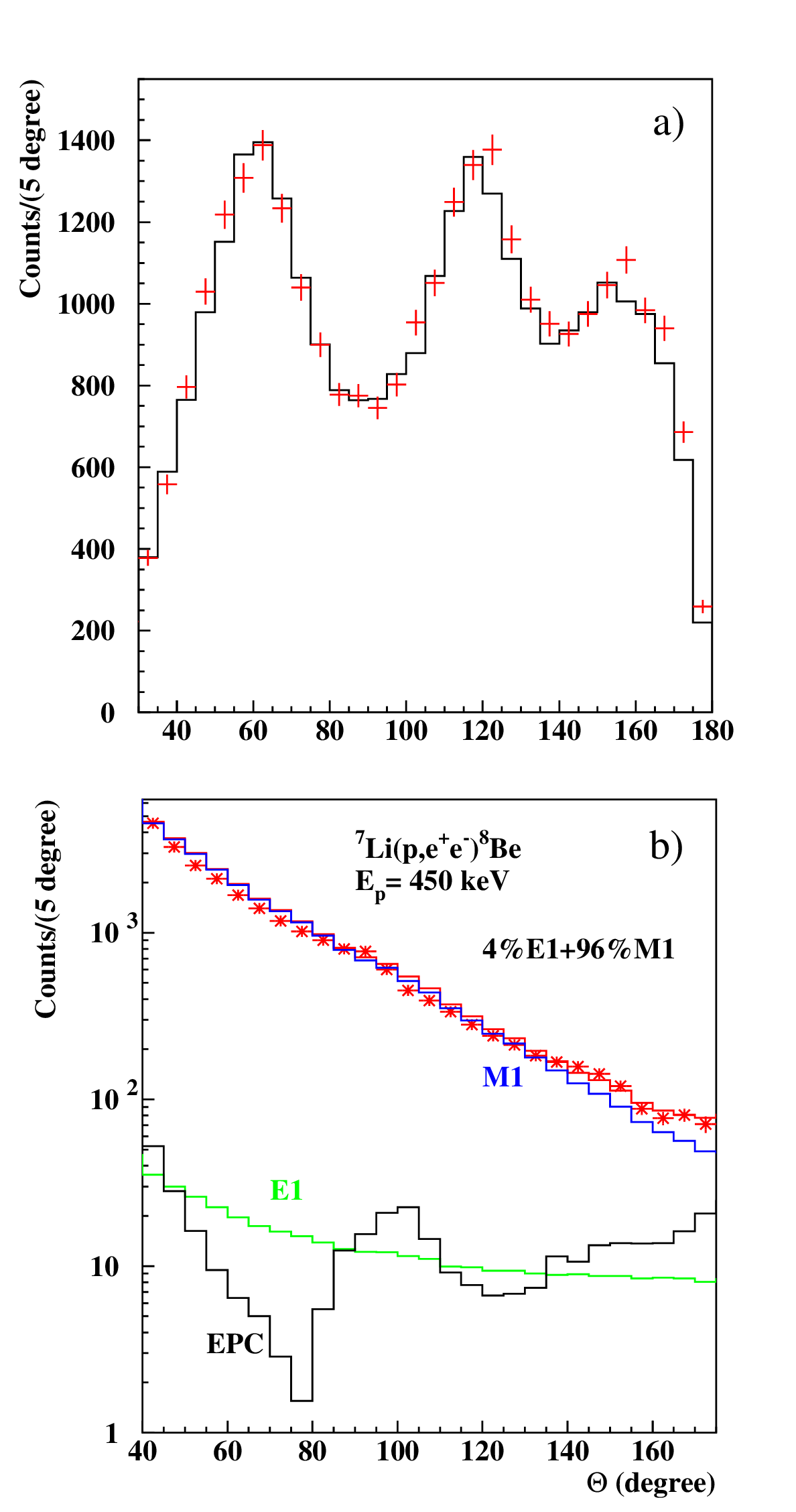}
  \end{center}
  \vspace{-0.5cm}
  \caption{a) Detector response for the setup as a function of
    correlation angle ($\theta$) for isotropic emission of e$^+$e$^-$
    pairs (red crosses) compared with the results of the Monte
    Carlo simulations (black line histogram) as explained in the
    text. b) $e^+e^-$ angular correlations obtained for the 17.6~MeV
    transition of $^8$Be by using thin target backing compared to the
    simulations performed for E1 and M1 IPC, as well as for the EPC created
    by the $\gamma$-rays on the different materials around the target.}
  \label{fig:lif-calibration1}
\end{figure}

Owing to the rather large level width ($\Gamma$ = 1.15 MeV
\cite{aj90}), a 2 mg/cm$^2$ thick $^{11}$B target was applied to
maximize the yield of the $e^+ e^-$ pairs. The target was evaporated
onto a 5 $\mu$m thick Ta foil. The average energy loss of the protons
in the target was $\approx$300 keV. To compensate for the energy
losses, the energy of the protons was chosen to be E$_p$ = 1.50, 1.70,
1.88, 2.10 and 2.50 MeV. The proton beam was impinged 
on a $^{11}$B target with a typical current of 2 $\mu$A 
for about 50 hours at each
beam energy. To achieve a more efficient cooling of the target and
thus to reduce its degradation, we replaced the previously used
\cite{kr16,gu16} Plexiglass support rods by Al rods. However, in turn,
our data suffered a bit larger signal background from Al induced by
the $\gamma$ rays as shown in Fig.~\ref{fig:lif-calibration1} than
before (Fig. 9 of Ref. [17]).

Our previous detector setup \cite{kr16,gu16} has recently been
upgraded. The details of the upgrade are described in \cite{kr21}.  
In the present experiment, the time and energy signals of the
scintillators, as well as the time,
energy and position signals of the DSSD detectors were recorded.

The energy calibration of the telescopes, the energy and position
calibrations of the DSSD detectors, the Monte Carlo (MC) simulations
as well as the acceptance calibration of the whole $e^+e^-$
coincidence pair spectrometer were explained in Ref. \cite{kr21}.
Reasonably good agreement was obtained between the experimental
acceptance and results of the MC simulations, as presented in
Fig.~\ref{fig:lif-calibration1}. The average difference is within
$\approx 3.0$\% in the angular range of 40$^\circ$~-~170$^\circ$.

In order to validate the accuracy of the MC simulations, we performed
measurements also on the $^{7}$Li($p$,$\gamma$)$^{8}$Be reaction.  The
experimental results for the angular correlations from this data
taking on the E$_p$~=~441~keV resonance (red dots with error bars) are
shown in Fig.~\ref{fig:lif-calibration1} b), together with the
corresponding Monte Carlo simulation (histogram) of the IPC process
stemming mostly from the M1 nuclear transition. The contribution of
the external pair creation (EPC) process of the 17.6~MeV $\gamma$-rays
is also shown by a black histogram.  We note here that the
direct-capture contribution is negligible compared to the M1 IPC due
to the large resonance capture cross section and the thin target.

As it can be seen in Fig.~\ref{fig:lif-calibration1}, the simulation of
the IPC process manages to describe the shape of the data
distribution accurately, and the contribution of EPC created on the
different parts of the spectrometer is reasonably low. 

In order to search for the assumed X17 particle, both the sum-energy
spectrum of the $e^+e^-$ pairs measured by the telescopes, and their
angular correlations, determined by the DSSD detectors, have
been analyzed.  Since the counting rates in the detectors were low
($\approx 150$ Hz in the scintillators and ($\approx 25$ Hz in the DSSD
detectors) and the coincidence time window was sharp ($\approx 10$ ns)
the effect of random coincidences was negligible. In the followings we
show only the real coincidence gated spectra.

In the GEANT simulations, both $e^+ e^-$ pairs generated by internal
pair creation in the target and the $e^+ e^-$- pairs generated by
external pair creation in the Ta backing were taken into account. A
more detailed description of the simulations can be found in
Ref. \cite{kr21}.

\section{Experimental results and discussion}

\begin{figure}[htb]
\begin{center}
\includegraphics[scale=0.40]{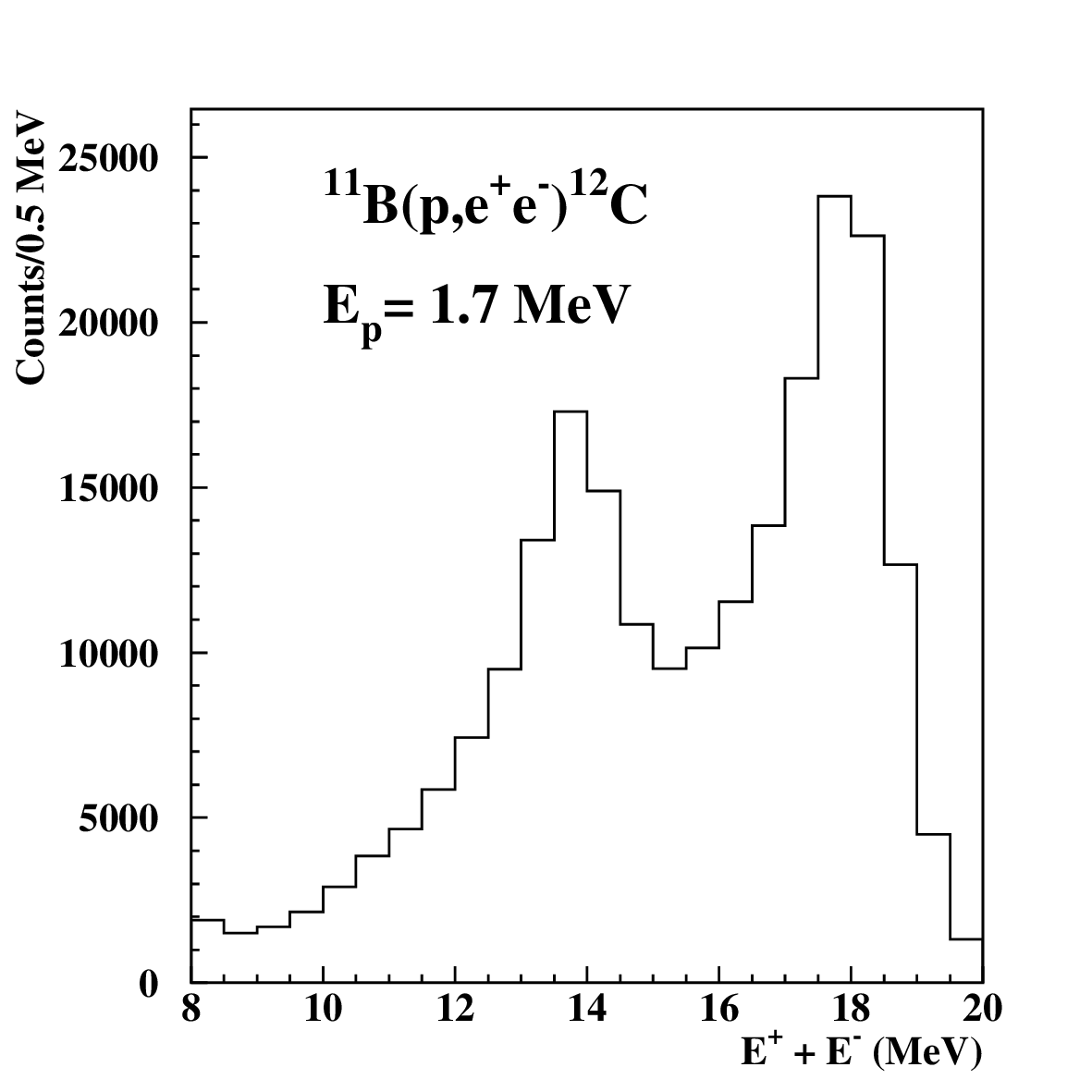}
\end{center}
\caption{Total energy spectrum of the $e^+e^-$-pairs from the
  $^{11}$B(p,$e^+e^-$)$^{12}$C nuclear reaction.}
\label{enesum}
\end{figure}

The total energy spectrum of the $e^+e^-$-pairs produced in the decay
of the E$_x$=17.2 MeV (J$^\pi$ = 1$^-$) state of $^{12}$C at E$_p$=1.7
MeV is presented in Fig.~\ref{enesum}. In addition to the E1 ground
state transition, this state decays to the E$_x$=4.44~MeV
(J$^\pi=2^+$) level with an E = 12.76 MeV E1 transition, which is also
present the energy spectrum. The intense, 4.44~MeV ground state
transition was discarded by setting a proper hardware threshold in
order to reduce the high count rate of the DAQ.

\begin{figure}[htb]
    \begin{center}
\includegraphics[scale=0.450]{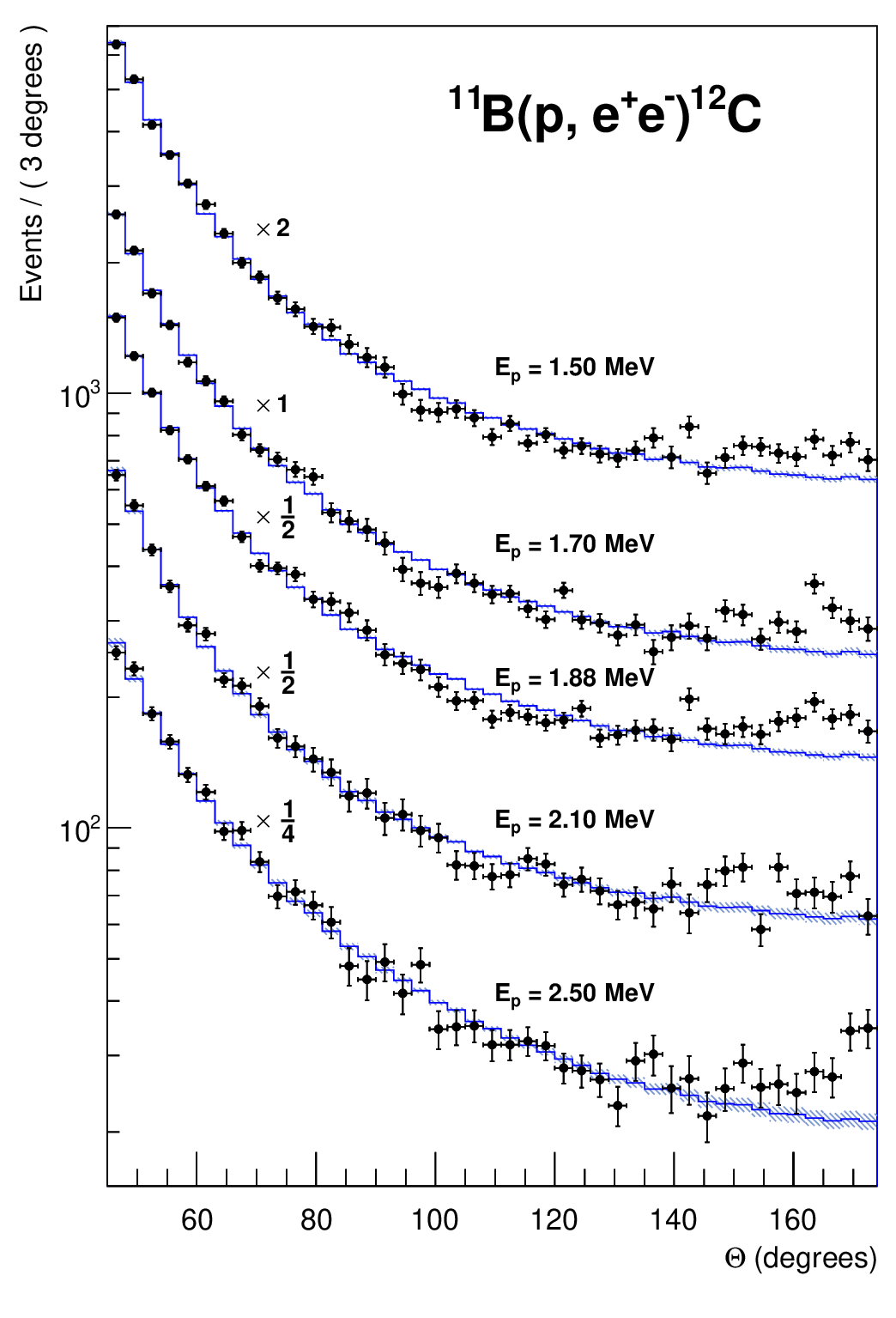}
    \end{center}
\caption{a) Experimental angular correlations of the $e^+e^-$ pairs measured in
  the $^{11}$B(p,$e^+e^-$)$^{12}$C reaction at the vicinity of the J$^\pi=1^+$ 
  resonance for different proton energies.}
     \label{ang-raw}
\end{figure}

The experimental efficiency of the $e^+e^-$ spectrometer was
determined with uncorrelated $e^+e^-$-pairs by taking the $e^-$ and
$e^+$ data from consecutive events as previously described in
Refs. \cite{kr16,kr17,kra17,kr21}. The gated and efficiency-corrected
$e^+e^-$ angular correlation in the 17.2 MeV (J$^\pi=1^-\rightarrow
0^+$) transition is shown in Fig.~\ref{ang-raw} for proton energies of
E$_p$= 1.5, 1.7, 1.88, 2.1 and 2.5 MeV.

As show in Fig.~\ref{ang-raw}, a combination of the MC simulated IPC
distributions of E1 and M1 radiations together with a small
contribution of simulated external pair creation (EPC) in the Ta
backing can describe the experimental distributions below $\Theta =
140^\circ$ reasonable well. However, we observe significant deviations
at large angles ($>140^\circ$) at each proton energy.

To derive the invariant mass of the decaying particle, we carried out
a fitting procedure for both the mass value and the amplitude of the
observed peak.

The fit was performed with RooFit \cite{Verkerke:2003ir} by describing
the $e^+e^-$ angular correlation with the following intensity function
(INT):
\begin{equation}
INT(e^+e^-) = N_{Bg} * PDF(exp) + N_{Sig} * PDF(sig)\ ,
\label{eq:pdf}
\end{equation}

\noindent
where $PDF(exp)$ was determined from a separate fit for the background
region, $PDF(sig)$ was simulated by GEANT for the two-body decay of
an X particle as a function of its mass, and $N_{Bg}$ and $N_{Sig}$
are the fitted numbers of background and signal events, respectively.

The signal PDF was constructed as a 2-dimensional model function of
the $e^+e^-$ opening angle and the mass of the simulated particle. To
construct the mass dependence, the PDF linearly interpolates the
$e^+e^-$ opening angle distributions simulated for discrete particle
masses.

\begin{figure}[htb]
\begin{center}
\includegraphics[scale=0.5]{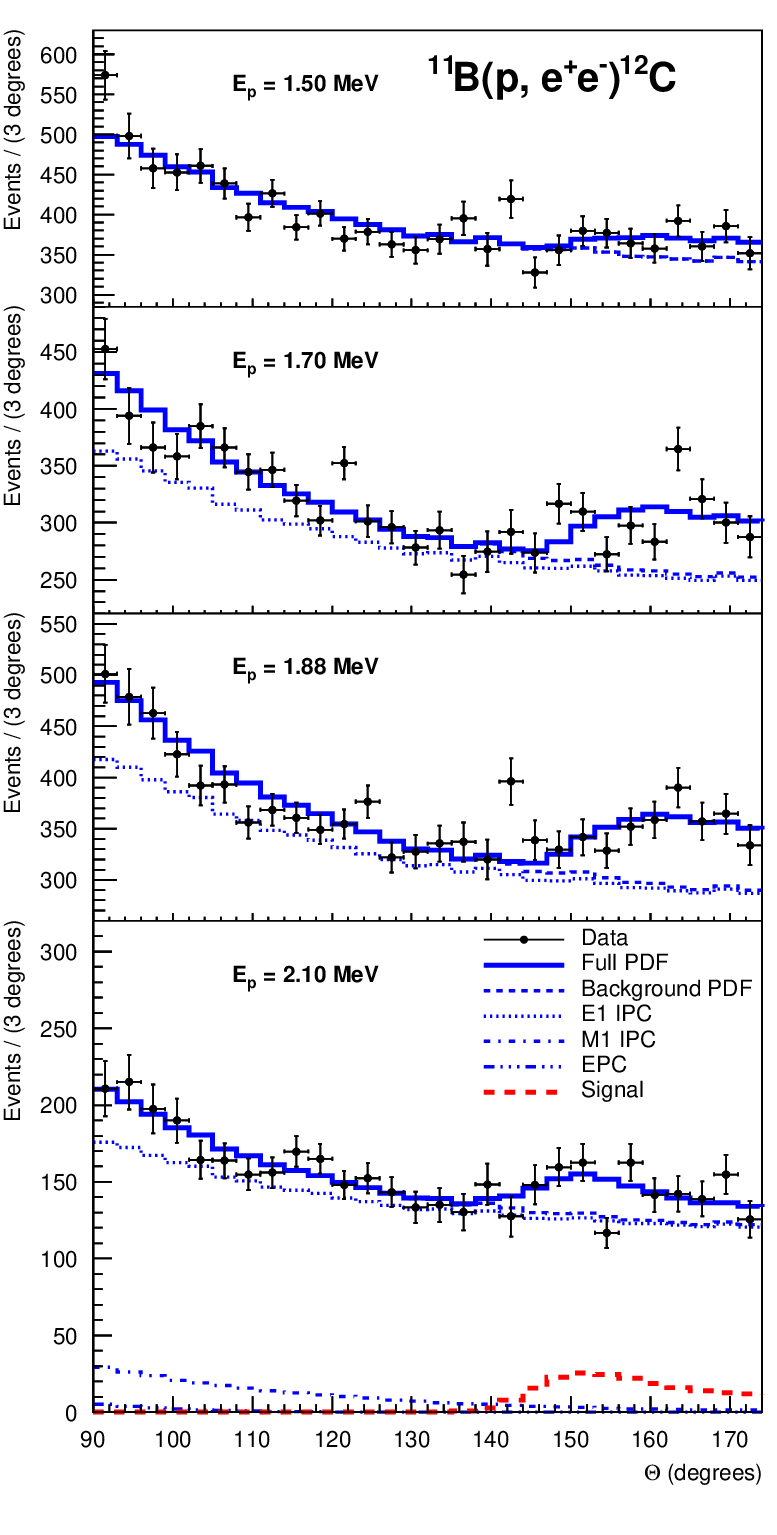}
\end{center}
\caption{Experimental angular correlations of the $e^+e^-$ pairs
  measured at different proton energies.  The full curves for each
  proton energy shows the results of the fit, using simulated angular
  distributions.}
 \label{bestfits}
\end{figure}

Using the intensity function described in Equation~\ref{eq:pdf}, we first
performed a list of fits by fixing the simulated particle mass in the
signal PDF to a certain value, and letting RooFit estimate the best
values for $N_{Sig}$ and $N_{Bg}$. Allowing the particle mass to vary 
in the fit, the best fitted mass is calculated 
and the corresponding fit is shown  for each studied beam energy in
Fig.~\ref{bestfits}.

A significant background is obtained from the E1 transition, but the
contribution from the assumed particle decay is also significant at
large angles.

The measured invariant masses of the hypothetical X17 and the
branching ratios (B$_x$) of its $e^+e^-$ decay to the $\gamma$ decay,
as derived from the fits are summarized in Table
\ref{tab:results}. The values are compatible for each fitted parameter
within 1$\sigma$ error bars. Their average values are also
highlighted.

\begin{table}[!h]
  \centering
  \caption{X17 branching ratios (B$_x$), masses, and confidences
    derived from the fits.}
  \label{tab:results}
  \begin{tabular}{llllc}
    \hline\hline
    E$_p$ & B$_x$           &Mass& Confidence \\\hline
    (MeV)&$\times10^{-6}$&(MeV/c$^2$)\\\hline
    1.50 & 1.1(6) & 16.81(15)& 3$\sigma$\\
    1.70 & 3.3(7) & 16.93(8)& 7$\sigma$\\
    1.88 & 3.9(7) & 17.13(10)& 8$\sigma$\\
    2.10 & 4.9(21) & 17.06(10)& 3$\sigma$\\\hline
    Averages&  3.6(3) & 17.03(11)&       &     \\\hline
    Previous \cite{kr16} &  5.8 & 16.70(30) &  &  \\\hline
    Previous \cite{kr21} &  5.1 & 16.94(12) &  &  \\\hline
    Predicted \cite{fe20} &  3.0 & & &      \\\hline\hline
  \end{tabular}
\end{table}

In Table~\ref{tab:results}, only the statistical errors are indicated.
The systematic uncertainties were estimated to be $\Delta
m_\mathrm{X}c^2$(syst.)~$= \pm 0.20$~MeV by employing a series of MC
simulations as presented in one of our previous works \cite{kr21}. It
mostly represents the uncertainty of the position of the beam spot,
which was found to be shifted by about $\pm 2$~mm in one measurement
run.

The extracted invariant mass agrees well with the values published
earlier for the $^8$Be \cite{kr16} and the $^4$He \cite{kr21}
experiments, which provides a convincing kinematic verification of the
existence of the X17 particle.  The branching ratio of the X17 decay
differs from the previous data, but, on the other hand, agrees well
with the theoretically predicted value \cite{fe20}.

\section{Summary}

We have studied the E1 ground state decay of the 17.2 MeV J$^\pi$ =
1$^-$ state in $^{12}$C. The energy-sum and the angular correlation of
the $e^+e^-$ pairs produced in the $^{11}$B($p$,e$^+$e$^-$)$^{12}$C
reaction were measured at proton energies of E$_p$= 1.50, 1.70, 1.88,
2.10 and 2.50 MeV. The gross features of the distributions of these
quantities can be described well by the IPC process following the
decay of the $1^-$ resonance and by considering a small contribution
of external pair creation process induced by the high-energy $\gamma$
rays. However, on top of the smooth, monotonic distribution of the
angular correlation of $e^+e^-$ pairs, we observed significant
peak-like anomalous excess around 155-160$^\circ$ at four different beam
energies.

The $e^+e^-$ excess can be well-described by the creation and
subsequent decay of the X17 particle, which we have recently suggested
\cite{kr16,kr21}. The invariant mass of the particle was derived to be
($m_\mathrm{X}c^2 = 17.03 \pm 0.11$(stat.)$\pm 0.20$(syst.)~MeV),
which agrees well with our previously published values. The branching
ratio of the $e^+e^-$ decay of X17 to the $\gamma$ decay was found to
be $3.6(3)\times10^{-6}$. The present observation of the X17 particle
in an E1 transition supports its vector character, as suggested by
Feng et al. \cite{fe20}.

Given the present results on the X17 creation in E1 transitions, we
consider to search for X17 in the decay of Giant Dipole Resonance
(GDR) excitations of different nuclei.

\section{Acknowledgements}

We wish to thank Z. Pintye for the mechanical and J. Molnar for the
electronic design of the experiment.  This work has been supported by
the Hungarian NKFI Foundation No.\, K124810 and by the
GINOP-2.3.3-15-2016-00034 and
\noindent GINOP-2.3.3-15-2016-00005 grants.

\end{document}